# QUANTIZED ARTIFICIAL NEURAL NETWORKS IMLEMENTED WITH SPINTRONIC STOCHASTIC COMPUTING

Saadi Sabyasachi[1], Walid Al Misba[1], Yixin Shao[2], Pedram Khalili Amiri[2] and Jayasimha Atulasimha[1, 3*]

[1] Department of Mechanical and Nuclear Engineering, Virginia Commonwealth University, Richmond, VA 23284, USA.
[2] Department of Electrical and Computer Engineering, Northwestern University, Evanston, IL 60208, USA.
[3] Department of Electrical and Computer Engineering, Virginia Commonwealth University, Richmond, VA 23284, USA.

*Abstract*—An Artificial Neural Network (ANN) inference involves matrix vector multiplications that require a very large number of multiply and accumulate operations, resulting in high energy cost and large device footprint. Stochastic computing (SC) offers a less resource-intensive ANN implementation with minimal accuracy loss. Random number generators (RNG) are required to implement SC in hardware. These can be realized through stochastic-magnetic tunnel junctions (s-MTJ), where the energy barrier to switch between the "up" and "down" states is designed to be small, enabling thermal noise to generate a random bit stream. While s-MTJs have previously been used to implement SC-ANNs, these studies have been limited to architectures with continuously varying (i.e., *analog*) weights. In this work, we study the use of SC for matrix vector multiplication with *quantized* synaptic weights and quantized outputs. We show that a quantized SC-ANN, implemented by using experimentally obtained s-MTJ bitstreams and using a limited number of discrete quantized states for both weights and hidden layer nodes in an ANN, can effectively reduce time (latency) and energy consumption in SC compared to an analog implementation, while largely preserving accuracy. We implemented quantization with 5 and 11 quantized states, along with SC configured with stochastic bitstream lengths of 100, 200, 300, 400, and 500 on neural networks with one hidden layer and three hidden layers. Inference was performed on the MNIST dataset for both training with SC and without SC. Training with SC provided better accuracy for all cases. For the shortest bitstream of 100 bits, the highest accuracies were 92% for one hidden layer and over 96% for three hidden layers. The overall system attained its peak accuracy of 96.82% using a 400-bit stochastic bitstream with three hidden layers. Our investigations demonstrate 9× improvement in latency to implement neuron activations and 2.6× improvement in energy consumption using the quantized SC approach compared to a similar s-MTJ based ANN architecture without quantization.

**\*Corresponding author: jatulasimha@vcu.edu**

## I. INTRODUCTION

Artificial Neural Networks (ANN) require a tremendous volume of computations which presents a challenge in implementing deep neural networks (DNNs) on edge devices where resources are at a premium. There has been substantial research in recent years [1], [2] to reduce the latency, power, and number of circuit components to implement efficient and reliable DNNs. One approach is to incorporate stochastic computing (SC) for inference on edge devices [3], [4]. In SC-based ANN, inputs, activations and tunable parameters such as weights are represented by a bitstream of "1"s and "0"s having a specific probability of observing "1". Matrix vector multiplications, the large computational loads of ANN, are replaced by probabilistic mathematics in SC, thus greatly reducing the hardware resources otherwise utilized in conventional computing based approaches [5], [6]. Several implementations of neural networks employing SC- multilayer perceptron (MLP) [7], [8], radial basis function NNs [9], convolutional neural networks (CNN) [10], deep belief networks (DBN) [11], and recurrent neural networks (RNN) [12] report competitive accuracy with significantly lower hardware cost and energy consumption [13]. Furthermore, SC remains an attractive choice for specific image processing tasks, such as, local image thresholding [14], low-precision real-time image processing [15] or kernel density approximation [16] due to faster and lower power implementation [17].

Efficient application of SC-ANN on edge devices relies on generating random numbers in an energy-efficient manner. Semiconductor (CMOS) circuits such as 32-bit linear feedback shift register (LFSR) can generate pseudo-random number which requires more than 1000 transistors making it resource-intensive [18]. In addition, non-volatile memory (NV) based technologies including phase change memory (PCM) [19], resistive random access memory (RRAM) [20] are being explored, along with the use of magneto resistive random access memory (MRAM) bits such as magnetic tunnel junctions (MTJs) [17]. MTJ is an attractive choice as it can generate stochastic bitstreams with compact design and low power [21]. Furthermore, MTJs are integrated with CMOS circuits in state-of-the-art semiconductor manufacturing foundries, making their states accessible with higher speed. The generated bitstreams can be tuned using techniques such as applying spin transfer torque (STT) [22] or spin–orbit torque (SOT) [23]. SC-based ANN using MTJ-generated bitstreams is reported in [24] for hand-written digit classification tasks. The stochastic bitstreams of continuous probabilities are generated using an analog approach [24].

However, to achieve ANN parameters (i.e. weights and biases) with analog precision (i.e. 32-bit) using non-volatile and other technologies often requires a lot of circuit overhead [25]-[27] and is more prone to device-to-device variations which can negatively impact the accuracy and scalability of the ANN [25]-[27]. Interestingly, ANN with extremely low precision or quantized weights and biases (can be even binary) are shown to achieve competitive accuracies [28]-[31] compared to 32-bit precision ANN. The accuracy degradation due to the quantization loss in such ANN is prevented during the learning stage by preserving the weight gradients [28]. The idea of quantization can be extended to SC where bitstreams with only a few probabilities are sufficient (i.e. 5 or 11-states) to retain the ANN accuracy. This way the peripheral circuit overhead and the complexity of the stochastic bitstream generating devices can be significantly reduced. Quantization-aware training using hardware-accelerators such as non-volatile computational devices arranged in crossbar architecture are reported with competitive accuracies [29]- [33], where matrix vector multiplications are performed by Kirchhoff's and Ohm's laws. In this paper, we explore the quantization aware training of SC-ANN where the inputs, activations and weights are represented by bitstreams of only a few quantized probabilities. The bitstream values with different probabilities are obtained using experiments where we bias the MTJs with different STT current densities [24]. During the quantization aware training, multiplication operations are simulated using bitwise XNOR operation (bipolar weights) and the addition operation with parallel counters. Furthermore, the sigmoid activation functions are implemented by using a much simpler look up table (LUT) based design due to the fact that the neuron output in quantized training can only assume a few discrete values (5 or 11 states). We tested the performance of our proposed training approach of the SC-based ANN on handwritten digit classification tasks using MNIST datasets where we evaluate the inference performance of such quantized SC-ANN on different topology ANNs (with 1 and 3 hidden layers) and explored the impact of smaller length bitstreams (such as 100-bits) on the accuracy. Using our approach with smaller size bitstreams (i.e. 100 bits length), we show that we can achieve competitive accuracy on MNIST handwritten digit classification tasks compared to 32-bit precision similar architecture ANN with much less energy consumption and latency. In addition, our approach requires low precision (i.e. 3-bit) digital to analog converter (DAC) which can potentially reduce prohibitive energy cost.

The rest of the paper is organized as follows. Section II provides a brief overview of stochastic computing implemented with random bit streams generated by s-MTJs, section III discusses the structure of deep neural networks and quantization, section IV presents a performance analysis and section V presents the conclusion.

## II. STOCHASTIC COMPUTING

### A. Probabilistic Mathematics

In stochastic computing (SC), a number is represented by the number of "1"s in its bitstream. Specifically, in unipolar encoding, the value of a number corresponds to the probability "1"s in its bitstream [34]. If a number X has N number of 1s and bitstream length is L, then stochastic representation of X will be $P(X)= N/L$. Multiplication of two numbers, $X=A \times B$ can be performed using simple AND gates, as in stochastic domain since $P(X)=P(A) \cdot P(B)$. Thus, the number of transistors is significantly reduced. Unipolar encoding can only represent positive numbers in the range [0, 1]. However, bipolar encoding is required to represent both positive and negative numbers in the range [-1, 1] [34]. In this scheme, a number is represented by $X= 2 \times P(X) – 1$ or alternatively, $P(X)= (X + 1)/ 2$.

To accommodate negative weights in the neural network, bipolar encoding has been used. XNOR gate can be used for multiplication of two bipolar numbers since $P(X)= P(A).P(B) + \overline{P(A)} \cdot \overline{P(B)}$ and for bipolar encoding it becomes $(X+1)/2 = [(A+1)/2] \times [(B+1)/2] + [1- (A+1)/2] \times [1- (B+1)/2]$ which produces $X=A.B$ [34]. Thus, the number of transistor gates required for arithmetic operations is also reduced significantly in this case.

### B. Bitstream Generation:

Using CMOS devices as shown in Figure 1(a), a random number generator can convert binary numbers to stochastic bitstreams. An alternative and more energy efficient way of generating Random Numbers is by using a MTJ shown in figure 2, consisting of two ferromagnetic layers separated by an oxide layer [24]. One layer has fixed magnetic orientation, referred to as the reference layer. The other layer's orientation can be switched. If the reference layer and free layer have magnetic orientation in the same direction, it corresponds to low resistance referred to as the Parallel (P) state ('0'). When the free layer has magnetic orientation in the opposite direction, it corresponds to higher resistance and is referred to as the Antiparallel (AP) state ('1') [35]. The two states are separated by an energy barrier $E_b$ and corresponding retention time $\tau = \tau_0 \exp\left(E_b/k_B T\right)$. Here $k_B$ denotes the Boltzmann constant and T denotes temperature. The energy barrier is determined by the interfacial magnetic anisotropy in the case of perpendicular magnetic tunnel junctions [36]-[38]. If $E_b$ is chosen to be small enough, then random thermally activated switching may occur at room temperature. For example, $E_b < 16 k_B T$ resulted in bitstream generation at a rate of milliseconds in [24], although much faster bit generation has been demonstrated in devices with smaller energy barriers or devices where the energy barrier is temporarily reduced by voltage controlled magnetic anisotropy (VCMA) [24], [39]-[41].

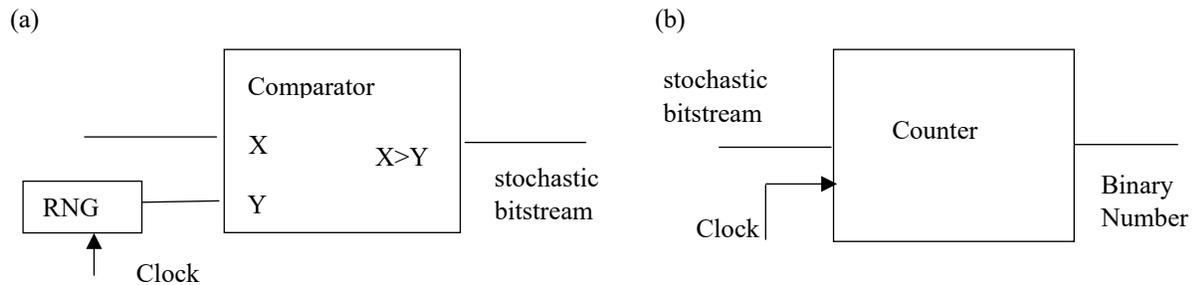

**Figure 1.** (a) Generation of stochastic bitstream, (b) Generation of Binary Number from stochastic bitstream [42]

In prior work, different bias voltages were applied to generate bitstreams corresponding to different input and weight values. It was shown that tunability from > 95% AP to > 95% P was achieved by bias voltages from -1V to +1V (figure 2c) [24]. About 30 different bias voltages were used to generate 30 different bitstreams per MTJ. The products (XNOR) of two MTJs were then utilized to increase resolution of each bitstream [24] and used for inference (not for learning with backpropagation) of a neural network for recognition of handwritten digits from MNIST dataset. However, this work [24] did not consider quantization of synaptic weights and outputs, which is implemented in our current work with only 5 or 11 quantized states. Such quantization could lead to significant improvements in latency and energy efficiency for inference, without significantly sacrificing inference accuracy.

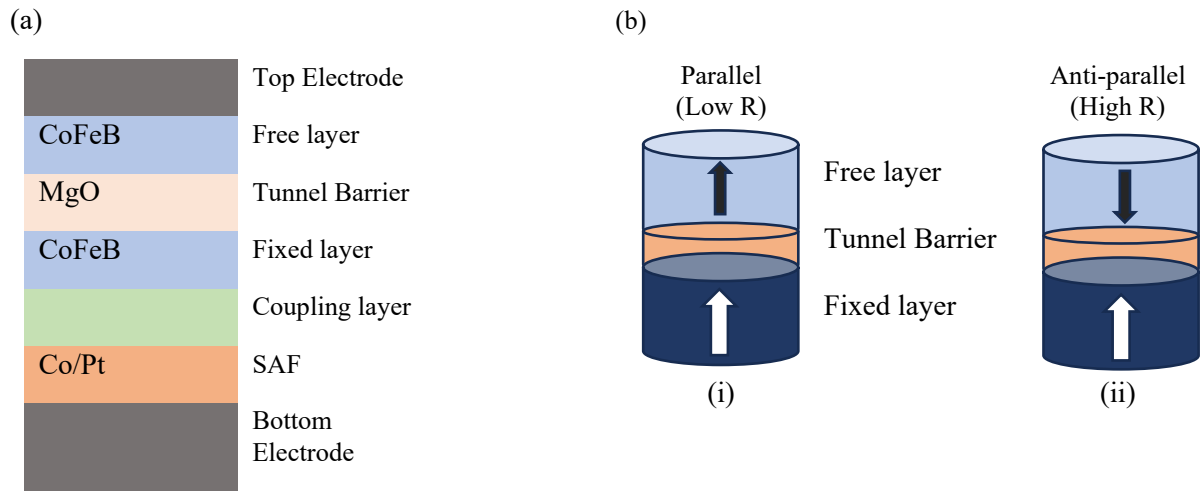

(c)

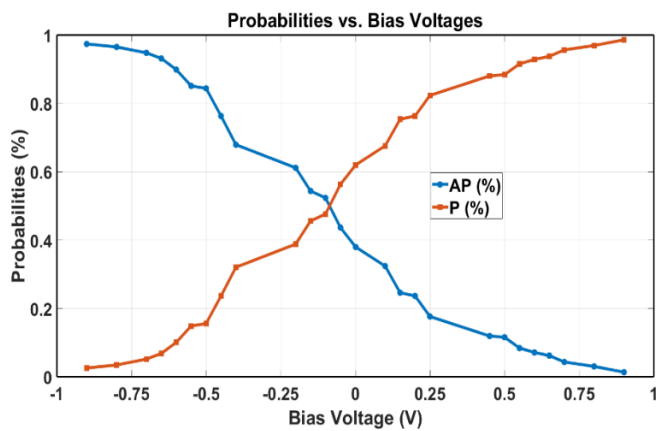

**Figure 2.** (a) Cross-section of MTJ [24], (b) Parallel and Anti-parallel states of MTJ [24], (c) Probabilities of 1s and 0s (parallel and antiparallel states) generated by an MTJ under different bias voltages [24].

### III. STRUCTURE OF DEEP NEURAL NETWORK AND QUANTIZATION OF SYNAPSES AND OUTPUTS

Two neural networks were implemented so the performance between the quantized and high precision implementation could be compared for a single hidden layer network and deeper network with three hidden layers. The input layer for both consists of 28×28 =784 pixels or data points. Architectures are summarized in Table I below. The network with one hidden layer is visualized in Figure 1.

Table 1: Architectures of one hidden layer and three hidden layer neural networks

Neural network-1: One hidden layer:

| Input layer nodes | Hidden layer-1 | Output layer nodes |
|---|---|---|
| 784 | 128 | 10 |

Neural network-2: Three hidden layers:

| Input layer | Hidden layer-1 | Hidden layer-2 | Hidden layer-3 | Output layer |
|---|---|---|---|---|
| 784 | 392 | 196 | 98 | 10 |

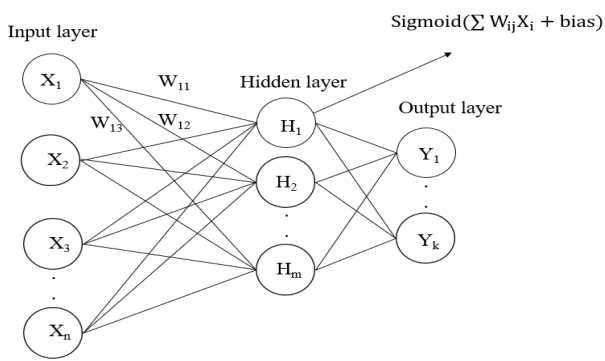

**Figure 3:** Neural network architecture with one hidden layer. The input data points are multiplied by their corresponding weights, followed by the application of a sigmoid activation function to generate the hidden layer nodes. Similarly, the output nodes are computed from the hidden layer. In our design, the network consists of 784 input neurons, 128 hidden neurons, and 10 output neurons (n = 784, m = 128, k = 10).

For the inference in the DNN, the multiplications were implemented using XNOR gates. For addition in binary domain, approximate parallel counter [43] or scaled multiplexer [44] can be used since SC-based networks generally allow some degree of precision loss. The activation function can be implemented for stochastic bitstream in various ways such as sigmoid and tanh functions implemented using JK Flip Flop or Finite State Machine (FSM) [45]. However, due to shorter bitstream and highly quantized (only 5 and 11 states) networks, parallel counter-based addition and look up table based sigmoid can be adopted without significant resource utilization, as explained in the following section.

**Quantization**: For practical implementation on edge devices, it would be resource intensive to generate all the probabilities. Hence, quantized weights have been used. In this network, weights have been scaled in the range of [-1, +1]. Weights or nodes are further quantized into 5 states or 11 states. We chose odd number of quantization states as they are centered around zero. Also, our past work [29], [31] showed that there is a significant gain in accuracy as one goes from 3-states to 5-states but gain is less significant beyond 5 states. Thus, the choice of 5-states was with an aim of achieving reasonable accuracy and the choice of 11-states was to see if there is a significant improvement in accuracy beyond 5-states. These quantized states are converted to corresponding stochastic bitstream for multiplication operation with bitwise XNOR. Afterwards, addition was performed with parallel counters. For implementation of the activation function, it is noticed that sigmoid activation function has only 3 or 6 distinct states (figure 4a) corresponding to 5 or 11 quantized states systems respectively. Hence sigmoid activation is implemented with look up tables.

After completion of feedforward pass, error is calculated from difference of predicted and actual output. Accordingly, weight gradients are calculated to update the weights. At the next training epoch, quantization of weights and inputs are implemented as before along to incorporate quantization aware training. Equations for quantization are presented below.

$$\text{Clip }(m, a, b) = \min\ (\max\ (m, a), b) \tag{1}$$

$$\Delta = \frac{b - a}{n - 1} \tag{2}$$

$$q = \left[\text{round}\left(\frac{\text{clip }(m, a, b) - a}{\Delta}\right)\right] \times \Delta + a \tag{3}$$

Here we have followed the approach of [29], [46], where n is the number of quantized states and a and b represent lower and upper limits.

Weights and nodes of the neural network are converted to corresponding stochastic bitstreams for multiplication operation with bitwise XNOR, after which addition is performed with a parallel counter. Sigmoid activation has only several distinct states so it can be implemented with look up tables (figure 4b).

**Learning:** For quantization-aware learning, methods of [29] were incorporated. Equations for learning are as follows.

$$\text{Cost function, } C = \frac{1}{2}\sum(y_i^L - d_i^L)^2 \tag{4}$$

$$\text{Gradient of cost function for output layer or error, } \delta_i^L = y_i^L - d_i^L \tag{5}$$

$$\text{Error for preceding layer, } \delta_i^l = W_{ij}\delta_j^{l+1} \tag{6}$$

$$\text{Weight update parameter, } \Delta W_{ij} = \eta x_i^l \delta_j^{l+1} f'_{l+1} \tag{7}$$

Here $\eta$= learning rate, $f'_{l+1}$= gradient of the activation function of layer l+1 neuron.

Here cost function is calculated by summation of square of error function defined by differences between predicted output and actual output (Eq. 4). Gradient of cost function for output layer is calculated by differentiation with respect to predicted outputs which can be termed as error (Eq 5). For preceding layers, errors were calculated by the backpropagation Eq. 6. Weights are updated as per Eq 7. Gradient of activation function is not included for back propagation though it is incorporated for weight updates [29]. It is to be noted that since differentiation for discrete or quantized values produce zero gradients, straight through estimator approach has been applied to backpropagate gradients [28], [29].

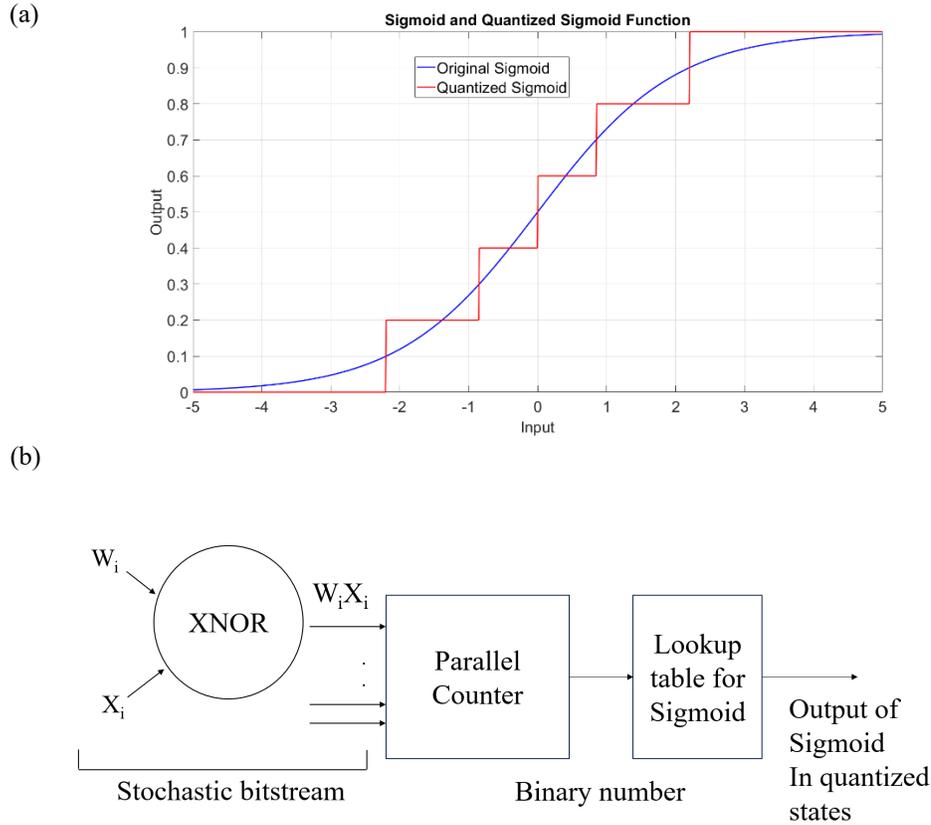

**Figure 4.** Implementation of the neural network. (a) Quantization of sigmoid activation function. (Even though 11 state quantization was used for the range [-1,1], output of sigmoid function only adopts 6 states since output of sigmoid is in the range [0,1]). (b) A simple schematic of our proposed quantized SC-ANN

**Algorithm: Quantization aware training for SC-ANN**

**Input:**
Training input dataset $X_0 = [X_1, X_2, X_3, \ldots X_N]$
Testing input dataset $T_0 = [T_1, T_2, T_3, \ldots T_M]$
// X, T are n dimensional vectors.
// number of layers= L, learning rate= $\eta$, QL= number of quantization states.

**Begin**

Weight initialization: W ← Gaussian distribution [-1, 1]
X ← X/ maximum (X);  // input data normalized;

// **feedforward**
**for** k= 1 to L **do**
    $X_{k-1}$ ← Quantize (Clip ($X_{k-1}$), −1, 1), QL)
    $W_k$ ← Quantize (Clip ($W_k$), −1, 1), QL)

```
        for i= 1 to num_neurons do

                X_k[i] = sigmoid ( ∑_j StochasticMult ( X_{k−1}[j], W_k [j, i]))  // j= number of inputs

        end for
        // stochastic multiplication implemented with XNOR for bipolar stochastic bitstreams
        // sigmoid activation implemented with look up tables.

end for

// Compute gradient
Compute gradient G_L= ∂C/∂X_L  from X_L and X_0

for k= L to 1 do
        G_{k-1} ← G_k W_k
        ΔW_k = ηG_k X_{k-1}
end for

// Update weight
for k= 1 to L do
        W_k (t+1) ← Update (W_k (t), ΔW_k)
        η ← λη   // λ denotes learning rate decay
end for
```

## IV. DISCUSSION AND ANALYSIS OF PERFORMANCE

As mentioned earlier, neural networks with one hidden layer and three hidden layers were tested with the MNIST dataset. Each network was tested for 5 quantized states and 11 quantized states for weights and hidden layer nodes. For stochastic Computing (SC), bitstream lengths of 100, 200, 300, 400 and 500 were used. Accuracy was determined for two scenarios for different bitstreams:

**Accuracy: i**. Method 1: Training was conducted for 5 and 11 quantized states across both networks without using SC. Inference computations were carried out both without SC and with SC, utilizing the mentioned bitstreams.

**ii.** Method 2: To explore the improvement of inference accuracy via SC training, the forward pass was carried out with SC and quantization while the backpropagation was carried out with high accuracy wights (not SC). Afterwards inference was conducted with and without SC, like the previous method.

The accuracy of the networks for different training and inference combinations are shown in figure 6.

For one hidden layer architecture, accuracy is always higher for training with SC. Besides, longer bitstreams generally provide higher accuracy. For 5 quantized states, the highest accuracy of 93.68% was achieved for 400 bits while for 11 states, highest accuracy of 94.48% was obtained for 500 bits.

For three hidden layer architecture, higher accuracy was also achieved for training with SC for all cases, similar to the one hidden layer architecture. However, the relationship between bitstream length and accuracy is not that straightforward for the deeper network. The highest accuracy obtained for 5 states is 96.62% for 300 bits and for 11 states it is 96.82% for 400 bits.  It can be theorized that noise arising from quantization and stochasticity errors assists the model in escaping the local minimum and moving closer to the global minimum of the loss function for the one hidden layer architecture. However, in a three-hidden-layer architecture, the cumulative effect of noise may become too large, potentially averaging itself out and failing to contribute effectively to approach the global minimum.

Besides, in the three hiddn layer architecture, the accuracy is constrained by number of quantized states. Hence it is unlikely that longer bitstream can attain any considerable improvement beyond 96.82% (highest attained accuracy for the systems). We also note that for a system where inference acuracy of 92% is sufficient, 100 bit SC system with one hidden layer architecture may be adequate while for a system with required accuracy of 96%, 100 bit SC system with three hidden layer architecture can be used.

The performance of these networks can be compared with other contemporary works. Reference [24] achieved 95% accuracy for 1024 bit long stream. In this work, 94.33% accuracy was obtained for 500 bits for 11 quantized weights for one hidden layer architecture. Daniels et al reported 97% accuracy using Lenet-5 architecture with bitstream length of 128 [25]. They used traditional digital circuitry to conrol bitstream statistics of SMTJ. Lenet-5 incorporates 6 hidden layers with multiple convolutional layers. Unipolar encoding for SC was incorporated, hence separate excitatory and inhibitory subnetworks were used to accommodate positive and negative weights. They used resolution of $\frac{1}{16}$ using SMTJ. Here we achieved similar accuracy with 100 bit long stream with 3 hidden layer architecture with broader resolution of $\frac{1}{2}$ and $\frac{1}{5}$ (5 and 11 quantized states).

(a)

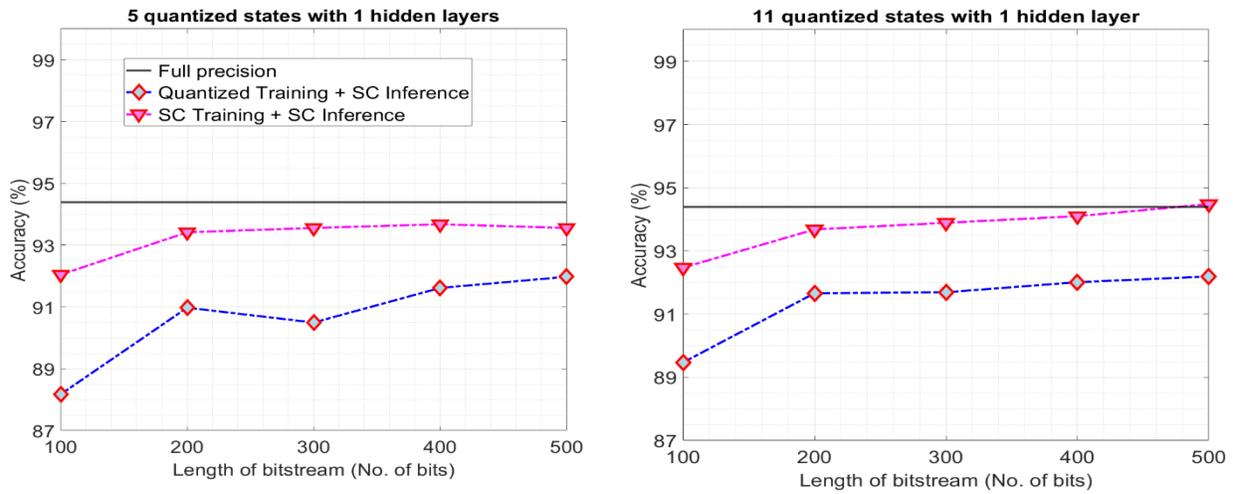

(b)

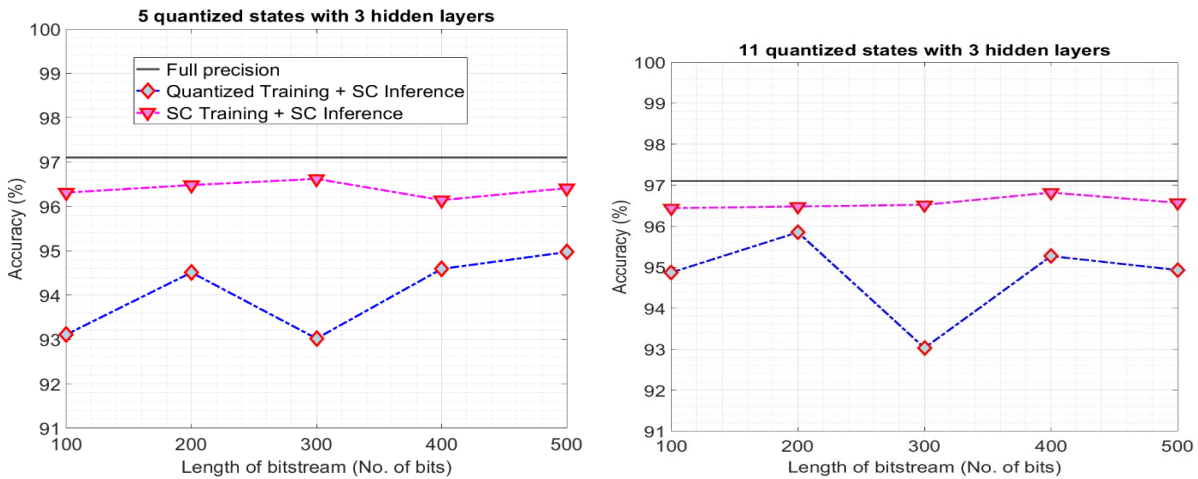

**Figure 6.** (a) Accuracy for one hidden layer with 5 and 11 quantized states. (b) Accuracy for three hidden layers with 5 and 11 quantized states.

One partiular distinction is that while we have incorporated quantization and SC for both training (forward pass) and inference, [24] and [25] used quantization and SC only for inference. SC inherently offers resilience to random bit flips to some extent, as a single random bit flip in the SC domain has less impact on the system compared to the binary domain, especially when the bit flip affects the most significant bit [47].

**Energy Consumption:**

Energy consumption for conventional CMOS based LFSR RNG per bit is ~10fJ [25]. For MTJ devices, energy dissipation is related to the retention time and the energy barrier $E_b$ between parallel and anti-parallel states. For a retention time of $\tau \approx 10$ ns, the energy per bit is ~20 fJ assuming an applied voltage of ~1 V and device resistance of 500 kΩ, which is comparable to CMOS-only RNGs [24]. With $\tau < 1$ ns, the energy can be reduced further [24].

We experimented with different bitstreams lengths. In [24] the authors reported 95% accuracy for 1024 bit long stream. We achieved 96.44% with 100 bits long stream with three hidden layers or 94.48% with 500 bits for one hidden layer. Thus, bitstream generation energy can be lowered from ~2100 nJ to 810~820 nJ for comparable accuracy with 2.6× reduction in energy. More importantly, our approach eliminates the requirement of digital to analog converter (DAC) which contributes to probably the biggest energy saving.

MTJ controlling circuitry also consumes energy. To generate bitstreams with deeper resolution in their respective probabilities, a complex combination of MTJ bitstreams and additional XNOR operations are required. In our proposed approach, bitstreams with only 5 or 11 different probabilities will suffice, which greatly simplifies the control circuitry and can be generated by biasing the MTJ with 5 or 11 different voltages.

Our quantized approach also comes with additional benefit in implementing the neuron activations such as sigmoid used in our works. As the neurons and weights assume only a smaller number of discretized states, the sigmoid can be implemented using a simple look-up-table (LUT), which will output only a few states. To test the latency improvement, we tested sigmoid activation implemented on FPGA using Xilinx software tools. A total of 9 clock cycle was needed to compute sigmoid using 11-state LUT while 80 clock cycle was required for similar 32-bit precision computation.

## V. CONCLUSION

SC-ANN with s-MTJ sourced stochastic bitstream has been tested for MNIST dataset with different bitstream lengths and quantized states. It was observed that even with shorter bitstreams of 100 bits and only 5 or 11 quantized states accuracy of 92% to over 96% was obtained. This approach achieved 9× improvement in energy and 2.6× improvement in latency, making it highly advantageous for neural network inference in edge computing applications.

## Acknowledgement

S.S., W.A. and J. A. acknowledge support from Department of Defense Airforce Grant #FA865123CA023 REQ F1TBAX3053A002 grant and National Science Foundation grant #1954589. The work at Northwestern was supported by the National Science Foundation under award numbers 1919109, 2322572, and 2311296.

**References:**

[1] J. Lee, S. Kang, J. Lee, D. Shin, D. Han and H. -J. Yoo, "The Hardware and Algorithm Co-Design for Energy-Efficient DNN Processor on Edge/Mobile Devices," in *IEEE Transactions on Circuits and Systems I: Regular Papers*, vol. 67, no. 10, pp. 3458-3470, Oct. 2020, doi: 10.1109/TCSI.2020.3021397.

[2] O. Spantidi, G. Zervakis, S. Alsalamin, I. Roman-Ballesteros, J. Henkel, and H. Amrouch, "Targeting DNN Inference Via Efficient Utilization of Heterogeneous Precision DNN Accelerators," IEEE Transactions on Emerging Topics in Computing, vol. 11, no. 1, pp. 112-125, Jan.-Mar. 2023, doi: 10.1109/TETC.2022.3178730.

[3] C. F. Frasser, P. Linares-Serrano, I. Díez de los Ríos, A. Morán, E. S. Skibinsky-Gitlin, and J. Font-Roselló, "Fully Parallel Stochastic Computing Hardware Implementation of Convolutional Neural Networks for Edge Computing Applications," IEEE Transactions on Neural Networks and Learning Systems, vol. 34, no. 12, pp. 10408-10418, Dec. 2023, doi: 10.1109/TNNLS.2022.3166799.

[4] M. Koo, G. Srinivasan, Y. Shim and K. Roy, "sBSNN: Stochastic-Bits Enabled Binary Spiking Neural Network With On-Chip Learning for Energy Efficient Neuromorphic Computing at the Edge," in *IEEE Transactions on Circuits and Systems I: Regular Papers*, vol. 67, no. 8, pp. 2546-2555, Aug. 2020, doi: 10.1109/TCSI.2020.2979826.

[5] K. Kim, J. Kim, J. Yu, J. Seo, J. Lee and K. Choi, "Dynamic energy-accuracy trade-off using stochastic computing in deep neural networks," 2016 53nd ACM/EDAC/IEEE Design Automation Conference (DAC), Austin, TX, USA, 2016, pp. 1-6, doi: 10.1145/2897937.2898011.

[6] J. Li, A. Ren, Z. Li, C. Ding, B. Yuan, Q. Qiu, and Y. Wang, "Towards acceleration of deep convolutional neural networks using stochastic computing," 2017 22nd Asia and South Pacific Design Automation Conference (ASP-DAC), Chiba, Japan, 2017, pp. 115-120, doi: 10.1109/ASPDAC.2017.7858306.

[7] J. L. Roselló, V. Canals and A. Morro, "Probabilistic-based neural network implementation," *The 2012 International Joint Conference on Neural Networks (IJCNN)*, Brisbane, QLD, Australia, 2012, pp. 1-7, doi: 10.1109/IJCNN.2012.6252807.

[8] Y. Liu, S. Liu, Y. Wang, F. Lombardi and J. Han, "A stochastic computational multi-layer perceptron with backward propagation", *IEEE Trans. Comput.*, vol. 67, no. 9, pp. 1273-1286, Sep. 2018.

[9] Y. Ji, F. Ran, C. Ma and D. J. Lilja, "A hardware implementation of a radial basis function neural network using stochastic logic", *Proc. Design Automat. Test Eur. Conf. Exhib. (DATE)*, pp. 880-883, 2015.

[10] Z. Li, A. Ren, J. Li, Q. Qiu, Y. Wang and B. Yuan, "DSCNN: Hardware-oriented optimization for stochastic computing based deep convolutional neural networks", *Proc. IEEE 34th Int. Conf. Comput. Design (ICCD)*, pp. 678-681, Oct. 2016.

[11] Y. Liu, Y. Wang, F. Lombardi and J. Han, "An energy-efficient online-learning stochastic computational deep belief network", *IEEE J. Emerg. Sel. Topics Circuits Syst.*, vol. 8, no. 3, pp. 454-465, Sep. 2018.

[12] Y. Liu, L. Liu, F. Lombardi and J. Han, "An energy-efficient and noise-tolerant recurrent neural network using stochastic computing", *IEEE Trans. Very Large Scale Integr. (VLSI) Syst.*, vol. 27, no. 9, pp. 2213-2221, Sep. 2019.

[13] Y. Liu, S. Liu, Y. Wang, F. Lombardi and J. Han, "A Survey of Stochastic Computing Neural Networks for Machine Learning Applications," in *IEEE Transactions on Neural Networks and Learning Systems*, vol. 32, no. 7, pp. 2809-2824, July 2021, doi: 10.1109/TNNLS.2020.3009047.

[14] M. H. Najafi and M. E. Salehi, "A fast fault-tolerant architecture for Sauvola local image thresholding algorithm using stochastic computing", IEEE Trans. Very Large Scale Integr. (VLSI) Syst., vol. 24, no. 2, pp. 808-812, Feb. 2016.

[15] A. Alaghi, C. Li and J. P. Hayes, "Stochastic circuits for real-time image-processing applications", *Proc. 50th Annu. Design Autom. Conf.*, pp. 1-6, 2013.

[16] P. Li and D. J. Lilja, "A low power fault-tolerant architecture for the kernel density estimation based image segmentation algorithm", *Proc. IEEE Int. Conf. Appl.-Specific Syst. Archit. Process*, pp. 161-168, Sep. 2011.

[17] B. R. Zink, Y. Lv and J. -P. Wang, "Review of Magnetic Tunnel Junctions for Stochastic Computing," in *IEEE Journal on Exploratory Solid-State Computational Devices and Circuits*, vol. 8, no. 2, pp. 173-184, Dec. 2022, doi: 10.1109/JXCDC.2022.3227062.

[18] Borders W A, Pervaiz A Z, Fukami S, Camsari K Y, Ohno H, Datta S (2019), "Integer factorization using stochastic magnetic tunnel junctions," Nature, vol. 573, pp. 390–393.

[19] E. Piccinini, R. Brunetti and M. Rudan, "Self-Heating Phase-Change Memory-Array Demonstrator for True Random Number Generation," in IEEE Transactions on Electron Devices, vol. 64, no. 5, pp. 2185-2192, May 2017, doi: 10.1109/TED.2017.2673867.

[20] J. Yang, Y. Lin, Y. Fu, X. Xue and B. A. Chen, "A small area and low power true random number generator using write speed variation of oxidebased RRAM for IoT security application," *2017 IEEE International Symposium on Circuits and Systems (ISCAS)*, Baltimore, MD, USA, 2017, pp. 1-4, doi: 10.1109/ISCAS.2017.8051019.

[21] E. Becle, G. Prenat, P. Talatchian, L. Anghel and I. -L. Prejbeanu, "A Fast, Energy Efficient and Tunable Magnetic Tunnel Junction Based Bitstream Generator for Stochastic Computing," in *IEEE Transactions on Circuits and Systems I: Regular Papers*, vol. 69, no. 8, pp. 3251-3259, Aug. 2022, doi: 10.1109/TCSI.2022.3173030.

[22] Y. Qu, J. Han, B. F. Cockburn, W. Pedrycz, Y. Zhang and W. Zhao, "A true random number generator based on parallel STT-MTJs," Design, Automation & Test in Europe Conference & Exhibition (DATE), 2017, Lausanne, Switzerland, 2017, pp. 606-609, doi: 10.23919/DATE.2017.7927058. [stt, trng]

[23] H. Chen et al., "Binary and Ternary True Random Number Generators Based on Spin Orbit Torque," 2018 IEEE International Electron Devices Meeting (IEDM), San Francisco, CA, USA, 2018, pp. 36.5.1-36.5.4, doi: 10.1109/IEDM.2018.8614638. [sot, rng]

[24] Y. Shao, S. L. Sinaga, I. O. Sunmola, A. S. Borland, M. J. Carey, J. A. Katin, V. Lopez-Dominguez, and P. Khalili Amiri, "Implementation of Artificial Neural Networks Using Magnetoresistive Random-Access Memory-Based Stochastic Computing Units," IEEE Magnetics Letters, vol. 12, pp. 1-4, 2021.

[25] M. W. Daniels, A. Madhavan, P. Talatchian, A. Mizrahi, and M. D. Stiles, "Energy-Efficient Stochastic Computing with Superparamagnetic Tunnel Junctions," Phys. Rev. Applied, vol. 13, no. 3, pp. 034016, Mar. 2020, doi: 10.1103/PhysRevApplied.13.034016.

[26] G. W. Burr, R. M. Shelby, A. Sebastian, S. Kim, S. Kim, S. Sidler, K. Virwani, M. Ishii, P. Narayanan, A. Fumarola, L. L. Sanches, I. Boybat, M. Le Gallo, K. Moon, J. Woo, H. Hwang, and Y. Leblebici, ``Neuromorphic computing using non-volatile memory,'' Adv. Phys., X, vol. 2, no. 1, pp. 89_124, Dec. 2016, doi: 10.1080/23746149.2016.1259585.

[27] G. W. Burr, R. M. Shelby, S. Sidler, C. Nolfo, J. Jang, I. Boybat, R. S. Shenoy, P. Narayanan, K. Virwani, E. U. Giacometti, B. N. Kurdi, and H. Hwang, "Experimental demonstration and tolerancing of a large-scale neural network (165 000 synapses) using phase-change memory as the synaptic weight element, " IEEE Trans. Electron Devices, vol. 62, no. 11, pp. 3498_3507, Nov. 2015, doi: 10.1109/TED.2015.2439635.

[28] I. Hubara, M. Courbariaux, D. Soudry, R. El-Yaniv, and Y. Bengio, "Quantized neural networks: Training neural networks with low precision weights and activations," *J. Mach. Learn. Res.*, vol. 18, no. 187, pp. 1_30, Apr. 2017.

[29] W. A. Misba, M. Lozano, D. Querlioz and J. Atulasimha, "Energy Efficient Learning With Low Resolution Stochastic Domain Wall Synapse for Deep Neural Networks," in *IEEE Access*, vol. 10, pp. 84946-84959, 2022, doi: 10.1109/ACCESS.2022.3196688.

[30] S. Dhull, W. A. Misba, A. Nisar, J. Atulasimha and B. K. Kaushik, "Quantized Magnetic Domain Wall Synapse for Efficient Deep Neural Networks," in *IEEE Transactions on Neural Networks and Learning Systems*, doi: 10.1109/TNNLS.2024.3369969.

[31] M. S. Alam, W. A. Misba, and J. Atulasimha, "Quantized non-volatile nanomagnetic domain wall synapse based autoencoder for efficient unsupervised network anomaly detection," *Neuromorphic Computing and Engineering*, vol. 4, no. 2, p. 024012, June 2024, doi: 10.1088/2634-4386/ad49ce.

[32] S. Liu, T. P. Xiao, C. Cui, J. A. C. Incorvia, C. H. Bennett, and M. J. Marinella, "A domain wall-magnetic tunnel junction artificial synapse with notched geometry for accurate and ef_cient training of deep neural networks," *Appl. Phys. Lett.*, vol. 118, May 2021, Art. no. 202405, doi: 10.1063/5.0046032.

[33] M. Ezzadeen, A. Majumdar, O. Valorge, N. Castellani, V. Gherman, G. Regis, B. Giraud, J.-P. Noel, V. Meli, M. Bocquet, F. Andrieu, D. Querlioz, and J.-M. Portal, "Implementation of binarized neural networks immune to device variation and voltage drop employing resistive random access memory bridges and capacitive neurons," Communications Engineering, vol. 3, p. 80, 2024, doi: 10.1038/s44172-024-00226-z.

[34] B. D. Brown and H. C. Card, "Stochastic neural computation. I. Computational elements," in IEEE Transactions on Computers, vol. 50, no. 9, pp. 891-905, Sept. 2001, doi: 10.1109/12.954505.

[35] Y. Zhang, X. Wang, Y. Li, A. K. Jones and Y. Chen, "Asymmetry of MTJ switching and its implication to STT-RAM designs," 2012 Design, Automation & Test in Europe Conference & Exhibition (DATE), Dresden, Germany, 2012, pp. 1313-1318, doi: 10.1109/DATE.2012.6176695.

[36] P. K. Amiri, Z. Zeng, J. Langer, H. Zhao, G. E. Rowlands, Y. Chen, I. N. Krivorotov, J. Wang, H. Jiang, J. A. Katine, Y. Huai, K. Galatsis, and K. L. Wang, "Switching current reduction using perpendicular anisotropy in CoFeB-MgO magnetic tunnel junctions," Applied Physics Letters, vol. 98, no. 11, p. 112507, 2011, doi: 10.1063/1.3567780.

[37] D. C. Worledge, G. Hu, D. W. Abraham, J. Z. Sun, P. L. Trouilloud, J. J. Nowak, S. L. Brown, M. C. Gaidis, E. O'Sullivan, and R. P. Robertazzi, "Spin torque switching of perpendicular Ta|CoFeB|MgO-based magnetic tunnel junctions," Applied Physics Letters, vol. 98, no. 2, p. 022501, 2011, doi: 10.1063/1.3542842.

[38] S. Ikeda, K. Miura, H. Yamamoto, K. Mizunuma, H. D. Gan, M. Endo, S. Kanai, J. Hayakawa, F. Matsukura, and H. Ohno, "A perpendicular-anisotropy CoFeB–MgO magnetic tunnel junction," Nature Materials, vol. 9, pp. 721–724, 2010, doi: 10.1038/nmat2804.

[39] Y. Shao, C. Duffee, E. Raimondo, N. Davila, V. Lopez-Dominguez, J. A. Katine, G. Finocchio, and P. Khalili Amiri, "Probabilistic computing with voltage-controlled dynamics in magnetic tunnel junctions," Nanotechnology, vol. 34, no. 49, p. 495203, 2023, doi: 10.1088/1361-6528/acf6c7.

[40] C. Duffee, J. Athas, Y. Shao, N. Davila Melendez, E. Raimondo, J. A. Katine, K. Y. Camsari, G. Finocchio, and P. Khalili Amiri, "Integrated probabilistic computer using voltage-controlled magnetic tunnel junctions as its entropy source," arXiv preprint, arXiv:2412.08017, 2024, doi: 10.48550/arXiv.2412.08017.

[41] C. Safranski, J. Kaiser, P. Trouilloud, P. Hashemi, G. Hu, and J. Z. Sun, "Demonstration of nanosecond operation in stochastic magnetic tunnel junctions," Nano Letters, vol. 21, no. 5, pp. 2040–2045, 2021, doi: 10.1021/acs.nanolett.0c04652.

[42] S. Liu, W. J. Gross and J. Han, "Introduction to Dynamic Stochastic Computing," in IEEE Circuits and Systems Magazine, vol. 20, no. 3, pp. 19-33, thirdquarter 2020, doi: 10.1109/MCAS.2020.3005483.

[43] K. Kim, J. Lee and K. Choi, "Approximate de-randomizer for stochastic circuits," 2015 International SoC Design Conference (ISOCC), Gyeongju, Korea (South), 2015, pp. 123-124, doi: 10.1109/ISOCC.2015.7401667.

[44] Sercan Aygun, M. Hassan Najafi, Lida Kouhalvandi, and Ece Olcay Gunes. 2024. Multiplexer Optimization for Adders in Stochastic Computing. In Proceedings of the 18th ACM International Symposium on Nanoscale Architectures (NANOARCH '23). Association for Computing Machinery, New York, NY, USA, Article 18, 1–2. https://doi.org/10.1145/3611315.3633256.

[45] K. K. Parhi and Y. Liu, "Computing Arithmetic Functions Using Stochastic Logic by Series Expansion," in IEEE Transactions on Emerging Topics in Computing, vol. 7, no. 1, pp. 44-59, 1 Jan.-March 2019, doi: 10.1109/TETC.2016.2618750.

[46] B. Jacob, S. Kligys, B. Chen, M. Zhu, M. Tang, A. Howard, H. Adam, and D. Kalenichenko, "Quantization and training of neural networks for efficient integer-arithmetic-only inference," 2017, arXiv:1712.05877.

[47] M. H. Sadi and A. Mahani, "Accelerating Deep Convolutional Neural Network base on stochastic computing," Integration, vol. 76, pp. 113-121, 2021, doi: 10.1016/j.vlsi.2020.09.008.